\documentclass[journal]{IEEEtran}
\ifCLASSOPTIONcompsoc
  \usepackage[nocompress]{cite}
\else
  \usepackage{cite}
\fi

\usepackage{outline}
\usepackage{pmgraph}
\usepackage[normalem]{ulem}
\usepackage[utf8]{inputenc}
\usepackage{amssymb}
\usepackage{hyperref}
\usepackage{amsmath}
\usepackage{graphicx}
\usepackage{times}
\usepackage{xcolor}
\usepackage{xspace}
\usepackage[colorinlistoftodos]{todonotes} 
\usepackage{cite}
\usepackage{bm} 
\usepackage{url}

\usepackage{epstopdf}
\AppendGraphicsExtensions{.tiff}
\graphicspath{{fig/}} 

\usepackage{epsfig}
\usepackage{tikz}
\usetikzlibrary{spy}
\usepackage{algpseudocode}
\usepackage{algorithm}
\usepackage{mathrsfs}

\usepackage{nicefrac}       
\usepackage{booktabs}       

\usepackage{amsmath,graphicx,caption,mathtools,amssymb,amsthm}
\usepackage{graphicx}
\usepackage{multirow}
\usepackage{subcaption}
\usepackage{accents}

\newcommand{\add}[1] {\textcolor{black}{#1}} 

\long\def\comment#1{} 

\newcommand{\xmath}[1] {\ensuremath{#1}\xspace}
\newcommand{\blmath}[1] {\xmath{\bm{#1}}}

\newcommand{\kb}{{\blmath k}}

\newcommand{\qb}{{\blmath q}}
\newcommand{\rb}{{\blmath r}}
\renewcommand{\sb}{{\blmath s}}
\newcommand{\tb}{{\blmath t}}

\newcommand{\xb}{{\blmath x}}
\newcommand{\yb}{{\blmath y}}

\newcommand{\Rc}{\mathcal{R}}

\newcommand{\Xc}{\mathcal{X}}

\newcommand{\Yc}{\mathcal{Y}}

\newcommand{\Rd}{{\mathbb R}}

\newcommand{\thetab}{{\boldsymbol {\theta}}}

\newcommand{\Ed}{{{\mathbb E}}}

\newcommand{\beq}{\begin{equation}}
\newcommand{\eeq}{\end{equation}}
\newcommand{\beqa}{\begin{eqnarray}}
\newcommand{\eeqa}{\end{eqnarray}}

\usepackage[utf8]{inputenc}
\usepackage{hyperref}
\hypersetup{
colorlinks=true,
linkcolor=blue,
filecolor=magenta,
urlcolor=cyan,
}
\usepackage{adjustbox}

\newcommand{\omegab}{\bm{\omega}}

\begin{document}
\title{Missing Cone Artifact Removal in ODT Using Unsupervised Deep Learning in the Projection Domain}
%
%
%

\author{Hyungjin Chung$^*$,
Jaeyoung Huh$^*$,
Geon Kim,
Yong Keun Park,
and~Jong~Chul~Ye,~\IEEEmembership{Fellow,~IEEE}
\thanks{H. Chung, J. Huh, and J.C. Ye are with the Department of Bio and Brain Engineering, Korea Advanced Institute of Science and Technology (KAIST), Daejeon 34141, Republic of Korea.}
\thanks{G. Kim and Y.K. Park are with the Department of Physics, Korea Advanced Institute of Science and Technology (KAIST), Daejeon 34141, Republic of Korea.}
\thanks{$^*$Equal contribution. }}

\maketitle

\begin{abstract}
Optical diffraction tomography (ODT) produces a three-dimensional distribution of the refractive index (RI) by measuring scattering fields at various angles. Although the distribution of the RI is highly informative, due to the missing cone problem stemming from the limited-angle acquisition of holograms, reconstructions have very poor resolution along the axial direction compared to the horizontal imaging plane. To solve this issue, we present a novel unsupervised deep learning framework that learns the probability distribution of missing projection views through an optimal transport-driven CycleGAN.
The experimental results show that missing cone artifacts in ODT data can be significantly resolved by the proposed method.
\end{abstract}


\IEEEpeerreviewmaketitle

\section{Introduction}
\IEEEPARstart{O}{ptical} diffraction tomography (ODT) is a technique to reconstruct a 3D refractive index (RI) distribution
\cite{lauer2002new,simon2008tomographic,wolf1969three,sung2009optical,jo2018quantitative,lee2013quantitative}.
Here, the refractive index (RI), $n$, is the optical property that relates to the electro- and magnetic susceptibility ($ \chi_e $ and $ \chi_m$, respectively) representing the whole light-matter interaction:
\begin{equation}
n(\tb) = \sqrt{[1+\chi_e(\tb)][1+\chi_m(\tb)]}, \quad \tb\in\Rd^3.\label{n-def}
\end{equation}
Since the light-matter interaction depends on the electron distribution and density, the RI has different values for each material. Thus, by reconstructing RI distributions, ODT provides not only the structure of cells but also their status, such as dry mass or chemical information\cite{popescu2006diffraction,popescu2004fourier,popescu2008optical,mir2011optical}.

\begin{figure*}[hbt!]
    \centering\includegraphics[width=16cm]{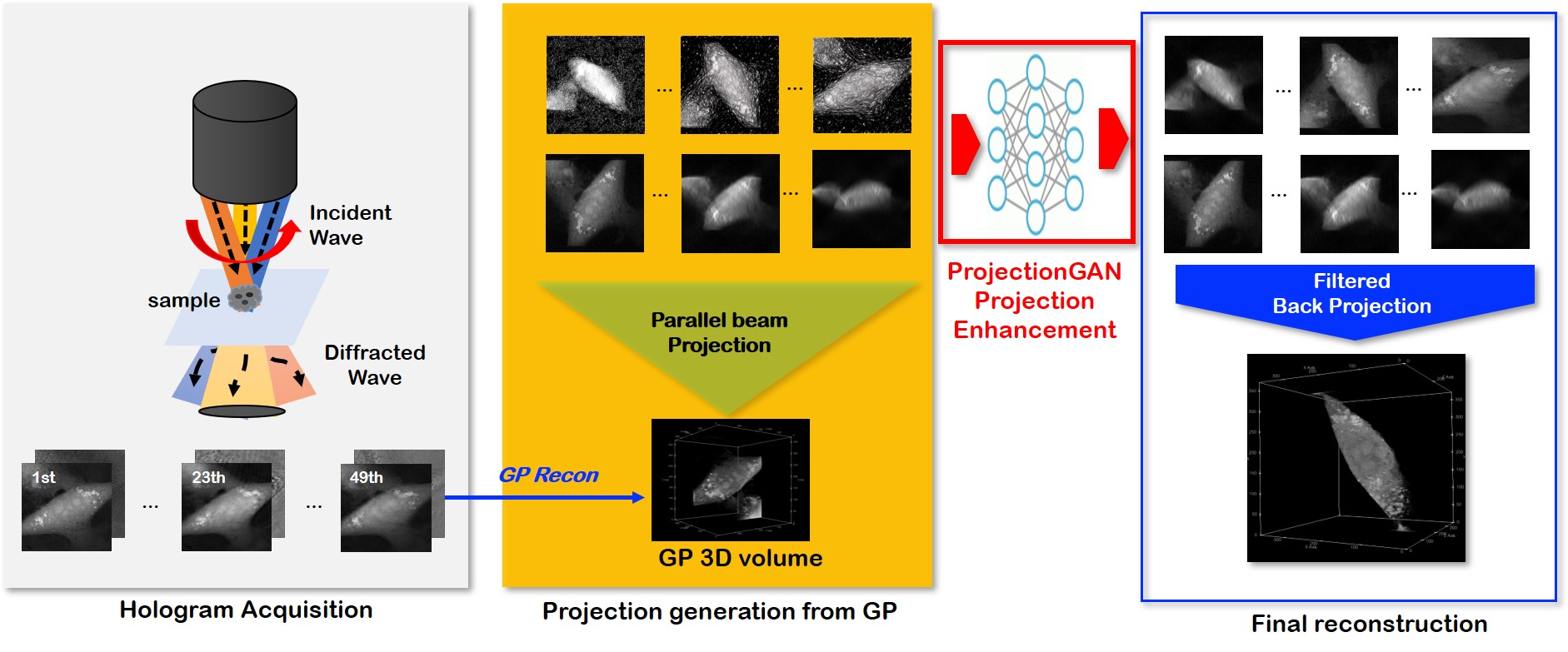}
\caption{Overview of the ProjectionGAN approach to resolve missing cone artifacts.
(Left) ODT measures diffracted waves in terms of holograms, in which acquisition angles are heavily limited.
(Center) The GP algorithm is applied for initial reconstruction. After reconstruction of the 3D RI data, parallel beam
projection was performed to generate 2D projection data.
(Right) Projection images are enhanced through the ProjectionGAN enhancement step,
after which final reconstruction is obtained with the FBP algorithm using the enhanced projection data.}
	\label{fig:fig1_overall_flow}
\end{figure*}

Despite its usefulness, technical challenges remain in ODT. In particular, resolution of the reconstructed RI distribution along the axial
direction is limited due to the well-known missing cone problem, which is a problem not specifically limited to the field of ODT and spans various fields involving tomographic imaging modalities~\cite{tam19793, tam1981tomographical, gerchberg1974super}. Typically, the 3D RI distribution can be obtained from multiple 2D holograms acquired from various illumination angles, as illustrated in Fig.~\ref{fig:fig1_overall_flow}. The waves become scattered when they encounter a medium, and such diffracted waves are measured as 2D optical fields called holograms. To acquire the 3D RI distribution from these holograms, measurements are mapped into the 3D Fourier space according to the Fourier diffraction theorem~\cite{kak2001principles}, after which follows the final inverse Fourier transform step. This approach results in missing Fourier components in cone-shaped angular regions, which introduces image distortions that make quantification of the RI values difficult. Furthermore, recent state-of-the-art systems \cite{shin2015active,lee2017time} acquire holograms using a circular sampling pattern with extremely sparse view sampling (e.g., 49 views). This is because the number of rotating mirror patterns that can be saved in the digital micromirror device (DMD) is limited. Moreover, typical imaging configurations of the current ODT system do not allow rotation of the samples that are being visualized \cite{shin2015active,lee2017time}. Due to these limitations, deficient projection views often yield technical problems, including more dominant missing cone artifacts, low-quality images with inaccurate RI values, etc., which makes the accurate quantification of the original RI values difficult. {Hence, devising a method for ODT reconstruction that eliminates missing cone artifacts is of great importance.}

Many researchers have investigated various regularization methods to address this problem\cite{tam1981limited,macias1988missing,barth1989approximation,kamilov2015learning} using iterative reconstruction approaches that incorporate a priori knowledge about the RI distribution.
These methods alternate between the image space and the Fourier space, imposing constraints on each domain. One of the most widely used algorithms is the Gerchberg-Papoulis (GP) algorithm \cite{gerchberg1974super, papoulis1975new}, which simply imposes nonnegativity on the RI values. This method removes the missing cone artifacts to a certain extent while being easy to implement \add{and hence is widely used}. Optimization algorithms derived from compressed-sensing (CS) theory have also been proposed, utilizing edge-preserving or sparsity promoting regularizations such as $\ell_1$ and total variation (TV) \cite{charbonnier1997deterministic, tian2011low, delaney1998globally, lim2015comparative}.
Although these iterative approaches enhance the resolution compared to the original reconstruction approach, they cannot overcome the inherent missing cone problem.
Moreover, smoothness constraints such as TV regularization often introduce cartoon-like artifacts in the reconstruction~\cite{lim2015comparative}.

Over the years, inspired by the recent success of deep learning approaches to solve ill-posed inverse problems, several authors have applied deep learning for ODT \cite{nguyen20183d, lim2020three, zhou2020diffraction,ryu2020deepregularizer}.
Most of the deep learning-based methods for ODT focus on supervised learning, where ground-truth data acquisition is feasible. Specifically, when artifact-free reference data are attainable, the network learns point-to-point mapping from the artifact-corrupted reconstruction to the matched reference data.
Nevertheless, in many interesting real-world situations in ODT, such ground truth is impossible to measure directly. Although some works \cite{nguyen20183d, lim2020three,ryu2020deepregularizer} utilized phantoms to train the neural network in a supervised fashion, they lack generality in practical situations. \add{For example, \cite{nguyen20183d} verified their method only with simulated phantom data. While \cite{lim2020three} demonstrated their method with in vivo data, the application is limited to a specific type of biological cell, namely, red blood cells, which are relatively simple to model and approximate with phantoms.} On the other hand, unsupervised learning methods for ODT reconstruction are still \add{at} the early stage of development. For instance, Zhou {\em et al.} \cite{zhou2020diffraction} developed a DIP\cite{ulyanov2018deep}-based method that requires hours of computation per experiment. Therefore, interest in developing \add{fast} unsupervised learning methods without matched ground-truth data is high.

Recently, several works have been proposed to train a generative model $G$ that acts as a stochastic sampler with respect to the latent space noise vector and unknown parametric measurement models \cite{bora2018ambientgan, gupta2020cryogan}. For example, if the measurement $\yb\in \Yc$ from the unknown input signal $\xb\in \Xc$ is given
by the forward mapping $\Rc_\thetab$ parameterized by the unknown parameter $\thetab \in \Theta$ such that
$\yb=\Rc_\theta \xb$, \add{then  the authors of AmbientGAN~\cite{bora2018ambientgan} propose adversarial training to learn the distribution of $\yb$ combined with parameterized random forward mapping $\Rc_\thetab$.}
The idea was further extended in cryoGAN~\cite{gupta2020cryogan}, where the authors approached the reconstruction of cryo-EM data from unknown projection view angles. However, the aim of these stochastic generative models was more toward the estimation of the unknown forward mapping, which is not the case for ODT where the projection view angles are known. Moreover, the stochastic generative model requires 3D reconstruction at each step of training, demanding a significant amount of time and computational resources for training.

In contrast to these approaches, one of the most important contributions of this work is to use unsupervised learning to significantly improve the projection views at the missing cone angles, after which missing cone artifacts can be reduced in a subsequent ODT reconstruction step using a standard reconstruction algorithm. Specifically, we use CycleGAN~\cite{zhu2017unpaired} to learn the optimal transport map that can convert the probability distribution of the blurry projections to that of the high-resolution projection images at the acquisition angles. \add{While distribution matching in the projection space can be regarded as denoising the projections in the general sense, we demonstrate that the use of OT-CycleGAN~\cite{sim2020optimal} is quite superior to classical denoising methods (e.g., BM3D~\cite{dabov2006image}, wavelet denoising~\cite{donoho1995noising}).}

Since we only process the 2D projection data at each step of the network training, the computational time for our method is significantly smaller than that of stochastic sampling approaches that directly reconstruct 3D volumes during each step of training. Despite the simplicity, extensive experiments using a numerical phantom,  {\em real} microbead data, and {\em in vivo} complex biological cells confirmed that our method can successfully address the long-standing missing cone problem in ODT.

\add{The rest of the paper is structured as follows. We provide the readers with relevant theoretical background of ODT measurement and reconstruction in Section~\ref{sec:background}. Our main contributions including the design of ProjectionGAN are summarized in Section~\ref{sec:main_contributions}. A detailed explanation of the implementation and experiments is given in Section~\ref{sec:methods}. The results of simulation and in vivo experiments are demonstrated in Section~\ref{sec:results}, followed by conclusions in Section~\ref{sec:conclusion}}

\section{Background}
\label{sec:background}
\subsection{Forward Measurement Model}
\label{background_forward}
In ODT, the forward diffraction measurement is modeled by the scalar Helmholtz equation:
\begin{align}
	\label{eq:Helmholtz}
	\Delta U_s(\tb) + \kappa^2U_s(\tb) = -  q(\tb)U(\tb), \quad\tb\in\Rd^3,
\end{align}
where $ \kappa = {2\pi}/{\overline{\lambda}}$ is the wavenumber with $\overline{\lambda}$ being the incident light wavelength in the free space, $U_s(\tb):= U(\tb)-U_0(\tb)$, $\tb\in\Rd^3$ is the scattered field from the scatterer with illumination-dependent total electric field $U$, background electric field $U_0$, and density $q$ denotes the normalized scattering potential.

The integral form representation of Eq. (\ref{eq:Helmholtz}) is the so-called \emph{Lippmann-Schwinger} integral equation \cite{born2013principles}:
\begin{eqnarray}
	\label{eq:solution}
	U_s (\tb) &=& \int  p(\tb') \mathrm{Gr}(\tb,\tb') U(\tb') d\tb',
\end{eqnarray}
where
$ \mathrm{Gr}(\tb,\tb') = {e^{i\kappa|\tb-\tb'|}}/{4\pi |\tb-\tb'|}$ is the outgoing free-space Green's function of the 3D Helmholtz equation.
Unfortunately, $U(\tb')$ is a function of $p(\cdot)$, which makes the integral equation in \eqref{eq:solution} nonlinear, so that
the direct inversion of $p(\tb)$ from $U_s(\tb)$ is difficult.
Accordingly, using the Born and Rytov approximation, \eqref{eq:solution} is linearized as
\begin{equation}
\begin{aligned}
\label{eq:Rytov}
	\phi_s (\tb) := &\Rc \qb(\tb) \\
	=  &\frac{1}{U_0(\tb)}\int  p(\tb') \mathrm{Gr}(\tb,\tb') U_0(\tb') d\tb',
\end{aligned}
\end{equation}

\begin{figure}[!t]
	\centering
\epsfig{figure=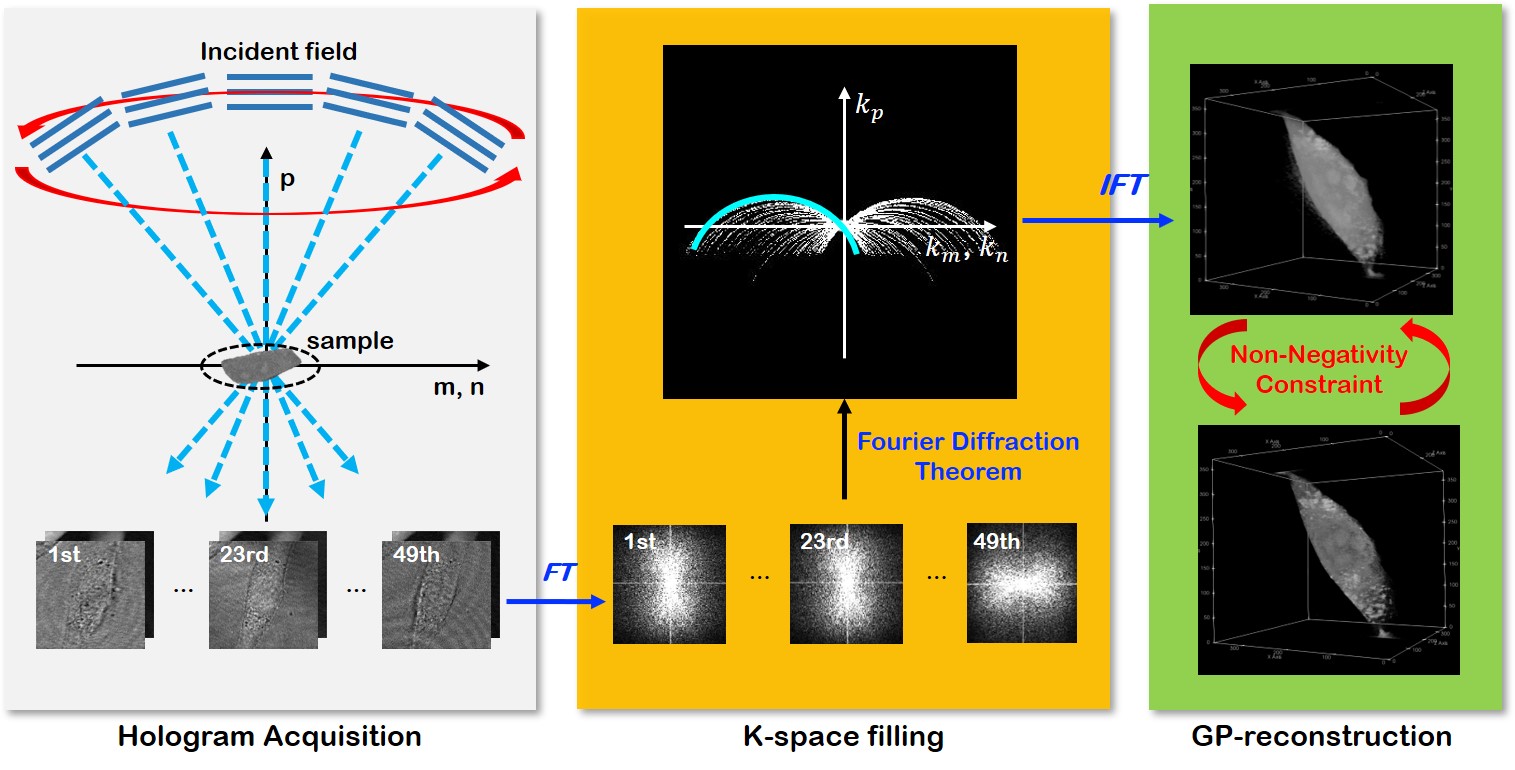, width=1.0\linewidth}
\caption{\bf\footnotesize
GP reconstruction flow in ODT. First, the phase object is illuminated with light, where the scattered illumination is measured in 2D holograms. Using Fourier diffraction theorem, appropriate 3D $k$-spaces are filled in. Then, the GP algorithm is applied, which incorporates a nonnegativity constraint.}
	\label{fig:supple_fig1_ODTgp}
\end{figure}
Now, suppose that the incident field $ U_0(\tb) $ is a plane wave, i.e., $ U_0^{(m)}(\tb) = e^{i\textbf{k}_m \cdot \tb} $, where $\textbf{k}_m = \kappa \hat{\sb}_m$ is a real-valued spatial wavenumber determined by the directional unit vector $\hat\sb_m = (\hat{s}_x^{(m)}, \hat{s}_y^{(m)}, \hat{s}_z^{(m)})$ for the $m$-th illumination.
Using the approximation \eqref{eq:Rytov}, the so-called measured holograms $p_m$ at the $m$-th illumination angle at $z = 0$ is defined as follows:
\begin{equation}
\begin{aligned}
\label{eq:measurement}
    &q_m =\\
    &\frac{\kappa\hat{s}_x^{(m)}}{2\pi i} \iint U_0^{(m)}(\tb)\ln{\frac{U(\tb)}{U_0^{(m)}(\tb)}}
    e^{-i\kappa(\hat{s}_x^{(m)}x + \hat{s}_x^{(m)}y)}dxdy.
\end{aligned}
\end{equation}
Further, we can formulate the forward operation $A_m p$ at position $(\kb - \kb_m)$, which reads
\begin{equation}
\label{eq:forward}
    A_m p = \int q(\tb') e^{-i(\kb - \kb_m)\rb'}d\rb'.
\end{equation}
According to the Fourier diffraction theorem \cite{wolf1969three}, for each illumination angle, ODT measures Fourier samples along a specific 3D half-spherical shell, as shown in Fig.~\ref{fig:supple_fig1_ODTgp}. Since the illumination vector $\sb_m$ is only limited to specific angles, this results in missing cone angles where no Fourier data are measured from the detectors. Moreover, with limited projection views, the sampling pattern cannot fill all the $k$-spaces near its sampling trajectories. This means that there is missing information between each angle, which further degrades the reconstruction quality.

\add{By inspecting the forward measurement model, we see that by filling in the missing information between angles, we would be able to boost the accuracy of reconstruction. This observation leads to the innovation of ProjectionGAN, which will be covered extensively in Section~\ref{sec:main_contributions}.}

\subsection{Reconstruction Methods}
The de facto standard for solving the reconstruction problem, which also suppresses the missing cone artifacts, is iteratively solving for the following equation~\cite{lim2015comparative}:
\begin{equation}
\label{eq:PLS}
    \min_p L(p), \, \text{such that} \, Ap = q,
\end{equation}
where $Ap = q$ is the data fidelity term between the reconstruction and the measurement, and $L(p)$ is the regularizer that drives $p$ to preferred solutions. Prior knowledge about the scattering potential, including properties such as nonnegativity and sparsity~\cite{lim2015comparative,gerchberg1974super,papoulis1975new}, is leveraged to design the regularizer, and we briefly review the two most widely used algorithms.

\subsubsection{Gerchberg-Papoulis reconstruction}

The GP model is one of the most popular algorithms that iteratively imposes nonnegativity, which ensures that the scattering potentials do not have a lower RI value than the background. \cite{lim2015comparative,gerchberg1974super,papoulis1975new}. In the case of the GP algorithm, $L$ is simply a unit function activated at the indices where $\text{Re}(p) > p_b$, with $\text{Re}$ extracting the real component of the signal, and $p_b$ is the RI background value. Now, \eqref{eq:PLS} can be solved via projection onto convex sets (POCS)~\cite{kus2019holographic}. Specifically, the GP algorithm iteratively solves for the following equations:
\begin{align}
\label{eq:iter_gp}
    &\hat{P}'_j = (1 - \beta\Omega)P_{j-1} + \beta q \\
    &p_j = A^H \hat{P}'_j \notag \\
    &p'_j = \sigma_{+}(\text{Re}(p_j)) \notag \\
    &\hat{P}_j = Ap'_j. \notag
\end{align}
where $\hat{P} = Ap$ defines the 3D $k$-space of the RI distribution, $\beta$ controls the contribution of data fidelity, $\Omega$ is a binary masking matrix holding value in the sampled indices, and $\sigma_{+}$ is the function imposing nonnegativity. While the GP algorithm is preferred due to its relatively low computational cost and ease of implementation, it reduces the missing cone artifacts to only a limited degree \cite{lim2015comparative,gerchberg1974super,papoulis1975new}.

\subsubsection{Total variation reconstruction}
\label{reconmethod_TV}
TV regularizes \eqref{eq:PLS} with prior knowledge that the signal should be piecewise smooth with a limited number of salient edges. Formally, the regularizer is defined as
\begin{equation}
\begin{aligned}
\label{eq:TV_loss}
	L(p) = \sqrt{(\triangledown _{x}p)^2 +(\triangledown _{y}p)^2+(\triangledown _{z}p)^2}.
\end{aligned}
\end{equation}
Then, \eqref{eq:PLS} can be minimized by transforming the equation into two subproblems through the split Bregman method~\cite{goldstein2009split}, where the update of $p$ is processed by solving the equation using conjugate gradient (CG) descent~\cite{lim2015comparative}. While TV is known to be highly effective when reconstructing data with piecewise constant properties, it is also well known that in the case of complex objects such as biological cells, imposing TV leads to cartoon-like artifacts that make the images appear blocky~\cite{dobson1996recovery, chan2000high, lim2015comparative}.

\section{Main Contributions}
\label{sec:main_contributions}

\subsection{\add{Projection Analysis}}
Recall that one of the most popular reconstruction algorithms for ODT is the Gerchberg-Papoulis (GP) algorithm, which iteratively imposes nonnegativity in the RI distribution to fill in the missing $k$-space information.
Nonetheless, we focus on the fact that the missing information between each angle is represented as low-resolution and artifact-contaminated projection images at the missing angles even after GP reconstruction. If the missing cone problem had not existed, then the projection images of all angles would share the same characteristics; that is, they would belong to the same distribution.
Therefore, if the projection images at the missing angles could be enhanced by a certain algorithm, one could expect that the resulting reconstruction would be improved as well.

One could think that this problem is solvable, similar to cryoGAN~\cite{gupta2020cryogan}, which would require fitting the 3D data to the GPU, randomly projecting the object at every weight update step, and updating the RI values of the 3D object directly from the gradients. However, this approach requires a vast amount of computational resources and time, so here, we propose a much simpler method that can be applied where we know the measurement angles.

\begin{figure}[!hbt]
    \centering
    \includegraphics[width=0.9\linewidth]{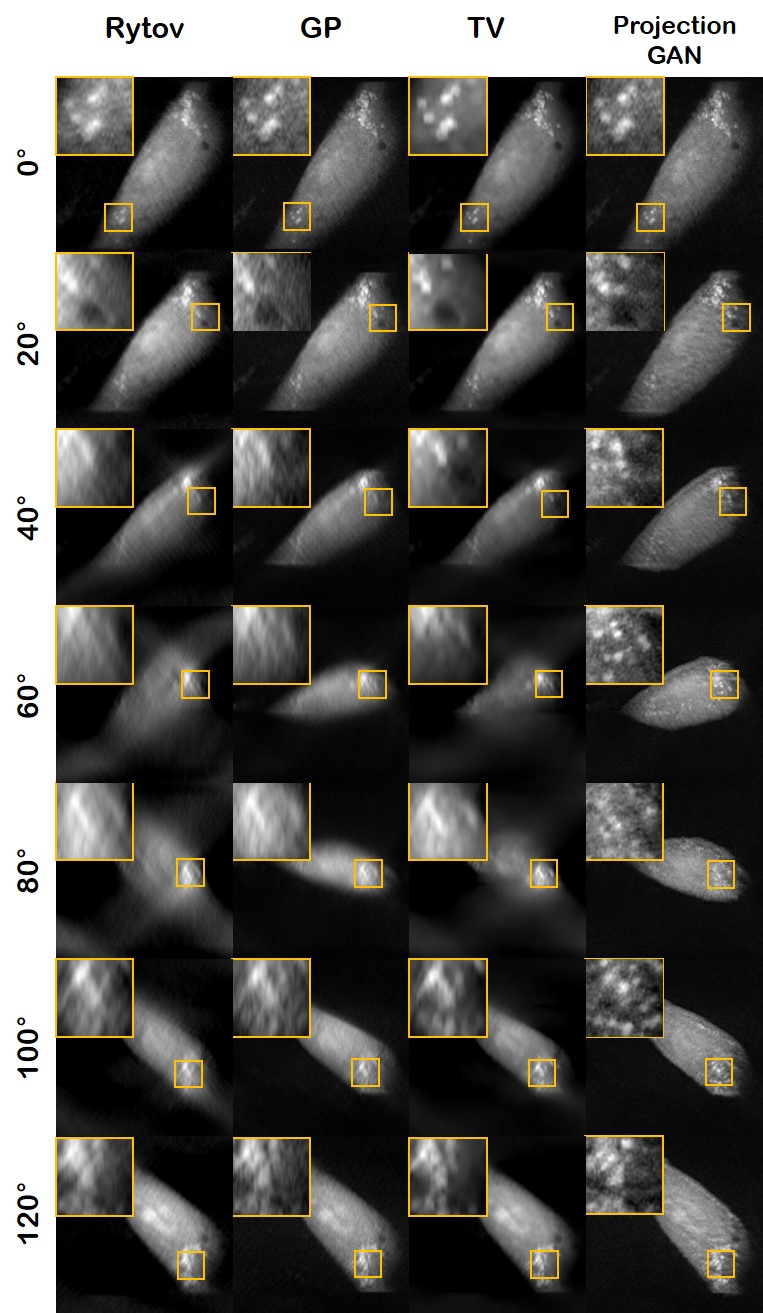}
\caption{Projection views of the NIH3T3 cell. {The projections were generated by rotating along the $y$-axis with the rotation angle given on the top. Each column corresponds to the reconstruction method. The resolution is degraded as the projection angle deviates from the measured angles} with all the other methods except ProjectionGAN. Resolution at all angles is well preserved with ProjectionGAN reconstruction.}
    \label{fig:projection_views}
\end{figure}

\subsection{\add{Fast Approximation with Projections}}
Once $k$-space filling through the Fourier diffraction theorem is done, the GP algorithm~\cite{gerchberg1974super, papoulis1975new} is used to initialize the volume. Starting here, we can {\em project} the 3D object into the 2D measurement space, as shown in Fig.~\ref{fig:fig1_overall_flow}. For projection, we use parallel beam projection. Although this is different from the actual forward model of the ODT, it still provides high-resolution projections at solid angles where the real holograms are measured, owing to the relation between the projection-slice theorem and projection-diffraction theorem~\cite{kak2001principles}
as shown in Fig.~\ref{fig:projection_views}.
Hence, we resort to parallel beam projection as a fast approximation. Note that by performing projection at the controlled angles, we can free ourselves from operating in the object space, which is in a prohibitively high dimension (e.g., in our case $372 \times 372 \times 372$) since the projection images can later be used for analytic reconstruction. Consequently, ProjectionGAN needs only to operate in the projection space, which is relatively easy to handle with low computational burden. This approach is vastly different from the prior AmbientGAN-like approaches~\cite{bora2018ambientgan,gupta2020cryogan}, where the forward projection operator would have to be applied to the massive 3D object per every gradient update step. Our approach only requires the projection step {\em once} prior to training.

\begin{table}[]
    \centering
    \begin{adjustbox}{width=0.25\textwidth}
    \begin{tabular}{c|c}
    \hline
         Method & Time (s) \\ \hline
         Rytov  & 1.37 \\ 
         GP     & 14.08 \\ 
         TV     & 604.86 \\ 
         CryoGAN~\cite{gupta2020cryogan} & Order of hours \\
         \textbf{ProjectionGAN} & \textbf{27.33} \\ 
    \hline
    \end{tabular}
     \end{adjustbox}
    \caption{Computation time required for reconstruction on an NVidia GeForce GTX 2080-Ti.}
    \label{tab:computation_time}
\end{table}

\add{The time required for reconstruction at the inference stage is compared in Table~\ref{tab:computation_time}. The ProjectionGAN enhancement step per sample takes approximately 13 seconds per sample, and with the GP initialization step added, it totals approximately 27 seconds. Hence, we can see that our method is much faster and computationally less expensive than modern iterative methods such as TV~\cite{lim2015comparative} or cryoGAN~\cite{gupta2020cryogan}.}

\subsection{\add{Unsupervised Distribution Matching}}

When the forward projection is performed, we achieve two projection image distributions: $\Yc_{\Omega}$ and $\Yc_{\Omega^c}$, where $\Omega$ denotes the solid angle for the {\em measured} projection angles, and $\Omega^c$ is the {\em unmeasured} or missing cone solid angles.
The projection images $\yb_{\omegab}$ at $\omegab\in \Omega$ usually have very high resolution without any missing cone angle artifacts,
whereas those of $\omegab \in \Omega^c$ suffer from blurring from the missing $k$-space information.
Therefore, our goal is to find the transform that improves the probability distribution on $\Yc_{\Omega^c}$ by learning that of $\Yc_{\Omega}$.

\begin{figure}[hbt!]
    \centering\includegraphics[width=8.5cm]{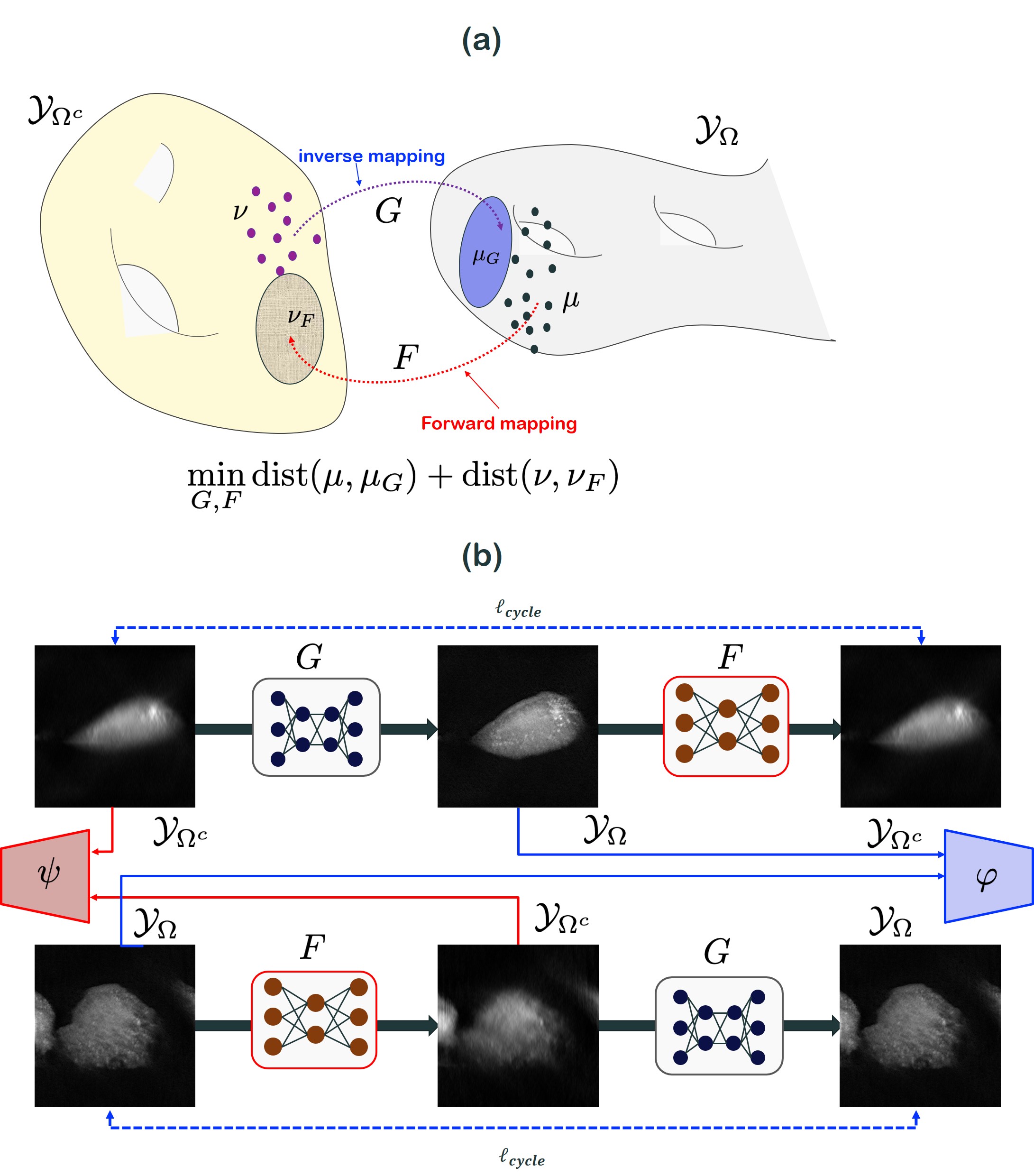}
\caption{(a) Geometric view of CycleGAN (b) The associated CycleGAN architecture.}
	\label{fig:ProjectionGAN}
\end{figure}

In fact, this problem can be rigorously addressed using the mathematical theory of optimal transport (OT).
More specifically, optimal transport is concerned with finding the optimal transport map that can transport one probability measure to another at the minimal transportation cost. Specifically, suppose that the target projection space $\Yc_\Omega$ is equipped with a probability measure $\mu$, whereas
the input projection space $\Yc_{\Omega^c}$ is \add{equipped} with a probability measure $\nu$, as shown in Fig.~\ref{fig:ProjectionGAN}(a).
Then, we can see that the mass transportation from the measure space $(\Yc_{\Omega^c},\nu)$ to another probability measure space $(\Yc_{\Omega},\mu)$ is done by a missing cone artifact-correcting generator $G$; i.e.,
the generator $G$ pushes forward the probability measure $\nu$ in $\Yc_{\Omega^c}$ to a measure $\mu_G$ in the target space $\Yc_{\Omega}$.
On the other hand, the mass transport from $(\Yc_{\Omega},\mu)$ to $(\Yc_\Omega^c,\nu)$ is performed by the artifact generation operator $F$
so that $F$ pushes forward the probability measure $\mu$ in $\Yc_{\Omega}$ to $\nu_F$ in the space $\Yc_{\Omega^c}$.
Then, the optimal transport map for unsupervised learning can be achieved by minimizing the statistical distances $\mathrm{dist}(\mu,\mu_G)$ between $\mu$ and $\mu_G$, and $\mathrm{dist}(\nu,\nu_F)$
between $\nu$ and $\nu_F$, and our proposal is to use the Wasserstein-1 metric as a means to measure the statistical distance.

In particular, if we minimize the two distances together using the Kantorovich OT formulation,
the following cycleGAN formulation can be derived \cite{sim2020optimal}:
\begin{align}
\min_{G,F}\max_{\psi,\varphi}\ell(G,F;\psi,\varphi)
\end{align}
with
\begin{align}\label{eq:loss_function}
\ell(G,F;\psi,\varphi):=  \lambda \ell_{cycle}(G,F) +\ell_{disc}(G,F;\psi,\varphi)
\end{align}
where $\lambda$ is an appropriate hyperparameter,
and the cycle-consistency loss is defined by
\begin{align*}\label{eq:cycleloss} 
\ell_{cycle}(G,F)  =& \int_{\Yc_{\Omega}} \|\yb- G(F(\yb)) \|  d\mu(\yb) \\
&+\int_{\Yc_{\Omega^c}} \|\yb'-F(G(\yb'))\|   d\nu(\yb') \notag 
\end{align*}
and the discriminator term is given by
\begin{align*}
&\ell_{disc}(G,F;\psi,\varphi)\\
=&\max_{\varphi} \int_{\Yc_{\Omega}}  \varphi(\yb)  d\mu(\yb) - \int_{\Yc_{\Omega^c}}  \varphi(G(\yb'))d\nu(\yb') \notag \\
 & + \max_{\psi} \int_{\Yc_{\Omega^c}}  \psi(\yb')  d\nu(\yb') - \int_{\Yc_{\Omega}}  \psi(F(\yb))  d\mu(\yb) \notag
\end{align*}
where $\varphi$ and $\psi$ are discriminator loss terms that
differentiate the real projections from the fake projections in $\Yc_\Omega$ and $\Yc_{\Omega^c}$, respectively. Here, the Kantorovich potentials should be 1-Lipschitz, and we use the LS-GAN formulation as our discriminator loss to impose finite Lipschitz conditions \cite{lim2020cyclegan}.

Once the projection views at the missing cone angle are improved, we use the standard 3D reconstruction algorithm for final reconstruction.
For the final 3D reconstruction from the improved parallel projection data, there could be many ways to acquire 3D tomography from projection datasets. For example, weighted backprojection (WBP)\cite{radermacher2007weighted}, Fourier reconstruction\cite{matej20013d}, and simultaneous iterative reconstruction (SIRT)\cite{trampert1990simultaneous} are commonly used in the cryo-EM community to account for projection reconstructions where the projection angle is random. As we can control the synthetic projection angles, we resort to a simpler method using filtered backprojection (FBP) with equal angular increments. More specifically, 360 equiangular projections are generated around each coordinate axis, which are obtained with a single generator $G$.
Subsequently, filtered backprojection (FBP) is applied to achieve the final reconstruction. To maintain the balance in all three axes, the three sets of reconstructed 3D potentials are averaged to obtain the final result.

\begin{figure}[!b]
\epsfig{figure=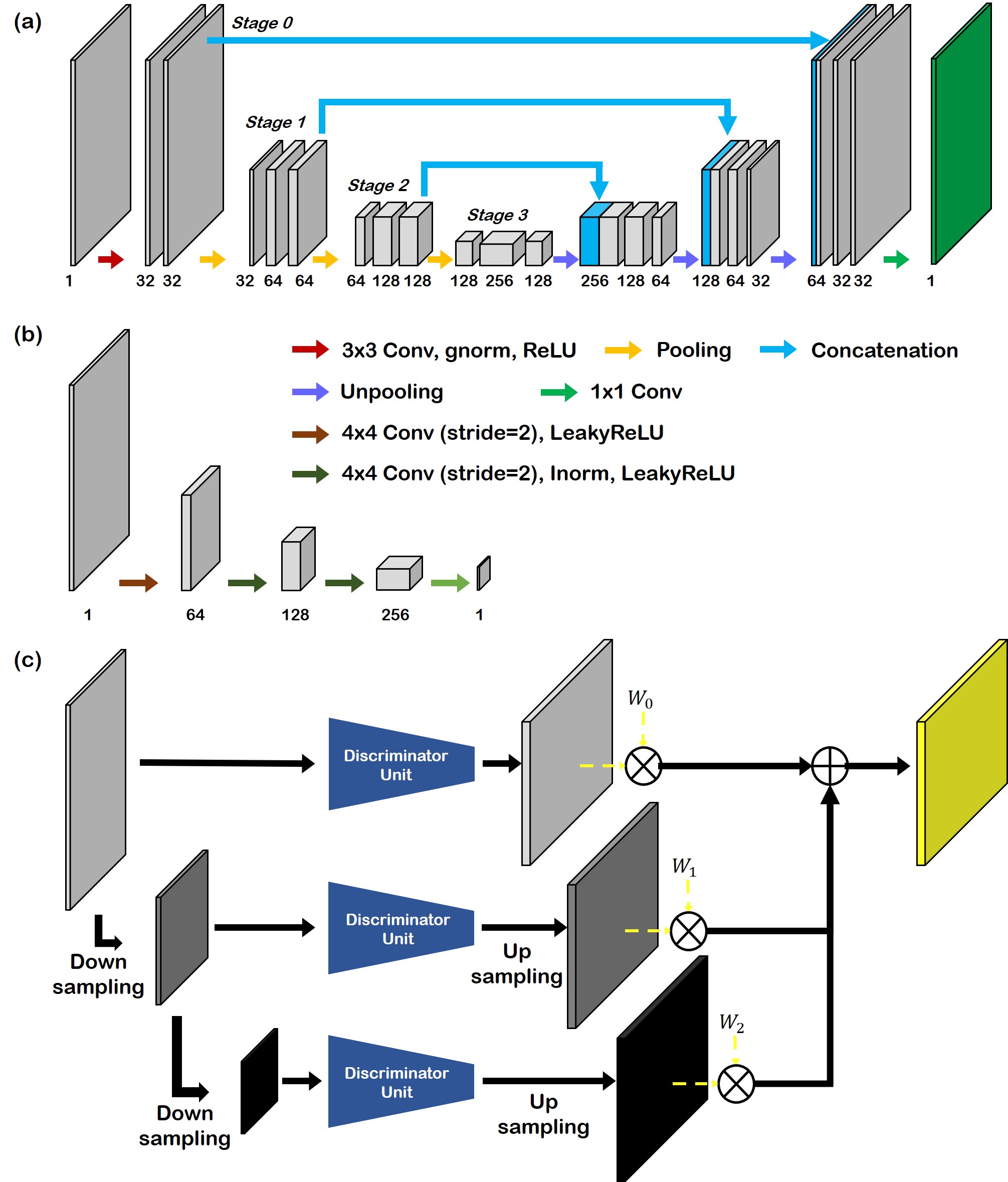, width=0.8\linewidth}
	\centering
\caption{\bf\footnotesize Neural network architecture used in the ProjectionGAN framework. (a) Generator architecture for $G$ and $F$, which is a slight modification of U-Net. (b) Discriminator unit, which was adopted from patchGAN of pix2pix. (c) Multiscale discriminator. Input is downscaled by a factor specified beforehand and passes through the discriminator unit, followed by upscaling and multiplication by the weight factor.}
	\label{fig:network_architecture}
\end{figure}

\section{\add{Experimental Setup}}
\label{sec:methods}

\subsection{Network Architecture}

The neural networks that were used for the training of CycleGAN are depicted in Fig.~\ref{fig:network_architecture}. The generator shown in Fig.~\ref{fig:network_architecture} (a) is an encoder-decoder architecture adopted from U-Net with a slight modification\cite{ronneberger2015u}. Specifically, the encoder and decoder parts were constructed with blocks of 3$\times$3 {convolution} layers, ReLU activation, and group normalization\cite{wu2018group}. For the pooling layer, we used a 3$\times$3 convolution layer with stride 2. For the unpooling layer, we used bilinear upscaling. Skip connections were used to concatenate features from each level of the encoder part, and a 1$\times$1 convolution layer was used at the final stage of reconstruction. For $G$, the initial number of filter channels in stage 1 was set to 32, which doubled per stage, reaching 256 channels at stage 3. For $F$, to limit the model capacity, we set the initial number of filter channels in stage 1 to 16, which doubled per stage, reaching 64 channels at stage 2.

The decoder architecture was adopted from patchGAN of pix2pix\cite{isola2017image} and is depicted in Fig.~\ref{fig:network_architecture} (b). All the convolution layers utilize 4$\times$4 convolutions with stride 2, which efficiently reduces the size of the image. LeakyReLU activation and instance normalization are used between the linear convolution layers. Finally, a 1$\times$1 convolution layer is placed at the end to produce prediction scores.

When multiple particles exist for reconstruction or when global characteristics and local characteristics differ, as in the case of the reconstruction of numerical phantom data, we found that using the multiscale discriminator shown in Fig.~\ref{fig:network_architecture} (c) and first introduced in Shocher {\em et al.}~\cite{Shocher2018InGANCA} is efficient. The architecture of the multiscale discriminator was adopted from Shocher {\em et al.}~\cite{Shocher2018InGANCA}, which has 4 different scale levels. The structure of the discriminator on each level shares the same architecture as in Fig.~\ref{fig:network_architecture} (b) without sharing the parameters at each scale. As the level progresses, the input data are downscaled by a factor of 2, with the lower levels of the discriminator focusing on a more fine-grained detail and the higher levels attending to the overall structure. The feature maps extracted on the higher levels are upscaled with bilinear interpolation and added to acquire the final output. The weights multiplied to each level are adjusted linearly, placing a larger emphasis on the details as the learning epoch progresses.

\subsection{Details of Training}

\subsubsection{Numerical Simulation}

For the training of numerical phantom data of size $372\times372\times372$, out of 27 sets of phantoms, 20 were used for training, and the other 7 were used for testing, which were generated by randomly placing spheres of different sizes. We set $\lambda = 0.01$ in Eq. \eqref{eq:loss_function}, with 4 levels of discriminator shown in Fig.~\ref{fig:network_architecture} (c). The weights applied to each level were set to $[1, 3, 5, 7]$ at the first epoch, which increased linearly up to $[7, 7, 7, 7]$ at the final epoch. Patch training was used with a size of 256$\times$256 sampled from a uniform distribution. Training was performed for 150 epochs, which took approximately one day.

Moreover, to conduct comparison studies with cryoGAN under our limited resources (NVidia GeForce GTX 2080-Ti: 11 GB RAM), we reiterated the same procedure but with a smaller size, $64\times64\times64$. For the experiment with this smaller size dataset, all configurations were kept the same, but the multiscale discriminator in Fig.~\ref{fig:network_architecture} was restricted to 3 levels by applying the weights $[1, 3, 5]$ at the first epoch, which linearly increased up to $[5, 5, 5]$ at the final epoch.

\subsubsection{ODT measurement data}

For training, out of 87 NIH3T3 cells and 18 microbead samples that were acquired, 67 different NIH3T3 cells and 10 microbead samples were selected randomly. The rest were utilized as test samples. Hyperparameter $\lambda$ for the loss function given in Eq. \ref{eq:loss_function} was set to 10 for the whole training process. Patch training was performed with a size of 256$\times$256 sampled from a uniform distribution. Training was performed for 20 epochs with a learning rate of 0.0001. Training took approximately one day.

With all experiments, the Adam optimizer was used with parameters $\beta_1 = 0.5$ and $\beta_2 = 0.999$~\cite{kingma2015adam}. The algorithm was implemented using PyTorch with NVidia GeForce GTX 2080-Ti as the GPU.

\subsection{Comparative algorithms}

To validate the efficacy of our algorithm, several existing methods are compared. The experimental details for each algorithm are as follows.
\subsubsection{GP}
We performed a standard GP procedure that repeatedly iterates between the k-space and the object space imposing a nonnegativity constraint, as expressed in \eqref{eq:iter_gp}. The parameters were set to $\beta = 1.0$, and 40 iterations were performed.

\subsubsection{TV}
For the optimization procedure for TV reconstruction, we followed~\cite{lim2015comparative}. However, we did not impose an additional nonnegativity constraint as did the authors of \cite{lim2015comparative}
because this provided better results in our experiments. We used $\mu = 900$ and $\alpha = 1000$ for the real ODT cell experiment and $\mu=100$ and $\alpha=100$ for the numerical simulation experiment. We used 5 iterations for CG optimization and 40 iterations for Bregman iteration.

\subsubsection{CryoGAN}
Although cryoGAN~\cite{gupta2020cryogan} was proposed for the reconstruction of cryo-EM data, by changing the forward model, we can also use cryoGAN for the reconstruction of ODT data. \add{In cryoGAN,
the measurement $\yb\in \Yc$ from the unknown input signal $\xb\in \Xc$ is given
by the forward mapping $\Rc_\theta$ parameterized by the unknown parameter $\theta \in \Theta$ such that
$\yb=\Rc_\theta \xb$. Then, one can utilize the following adversarial training to update $\xb$ directly:
\begin{equation*}
    \min_\xb \max_D \Ed_{\yb \sim P_\Yc} [D(\yb)] + \Ed_{\thetab\sim P_\Theta} [1 - D(\Rc_\theta(\xb))],
\end{equation*}
where $P_\Yc$ and $P_\Theta$ refer to the probability distribution for the measurement and the parameter space, respectively. $D$ denotes the discriminator that tells the fake measurements from the true measurements.}

\add{Specifically, we take parallel ray projection as $\Rc_\theta$, where $\theta$ indicates the solid angle at which the projection is performed.} In \cite{gupta2020cryogan}, the volume to be reconstructed is initialized with zeros; for fair comparison, we initialize the volume using the GP-reconstructed data. We keep $n_{discr} = 1$ and the number of training epochs to 1000, which took approximately 2 hours. Projection operator $\Rc_\theta$ \add{was implemented in} PyTorch using the open-source library torch-radon~\cite{torch_radon}.

\subsubsection{Projection-BM3D (Proj-BM3D)}
\label{sec:projection-BM3D}
One could also view ProjectionGAN as a denoiser operating in the projection domain. From this perspective, it is possible to apply any denoiser to the projections. To compare the efficacy of using cycleGAN for the denoiser specifically, we compared our approach with using BM3D~\cite{dabov2006image} as the projection domain denoiser. The BM3D results depend on the variance of noise, $v$, where higher values of $v$ make the structure more blurry. We utilized $v = 0.1$ in the experiment, which produced optimal results. The rest of the parameters were kept as advised in the original paper~\cite{dabov2006image}.

\subsubsection{\add{Projection-Wavelet Thresholding (Proj-wavelet)}}
\label{sec:projection-wavelet}
\add{Another widely recognized denoising method is denoising by wavelet thresholding~\cite{donoho1995noising, johnstone2004needles}. We implement wavelet denoising by first decomposing the image into 4 levels with the `Daubechies 3' \cite{daubechies1992ten} wavelet, hard thresholding with a value of 0.7, and combining the coefficients back to form the denoised image. The level of decomposition and the threshold value were chosen to produce optimal results.}

\section{Results}
\label{sec:results}

\begin{figure}[!hbt]
    \centering\includegraphics[width=8.5cm]{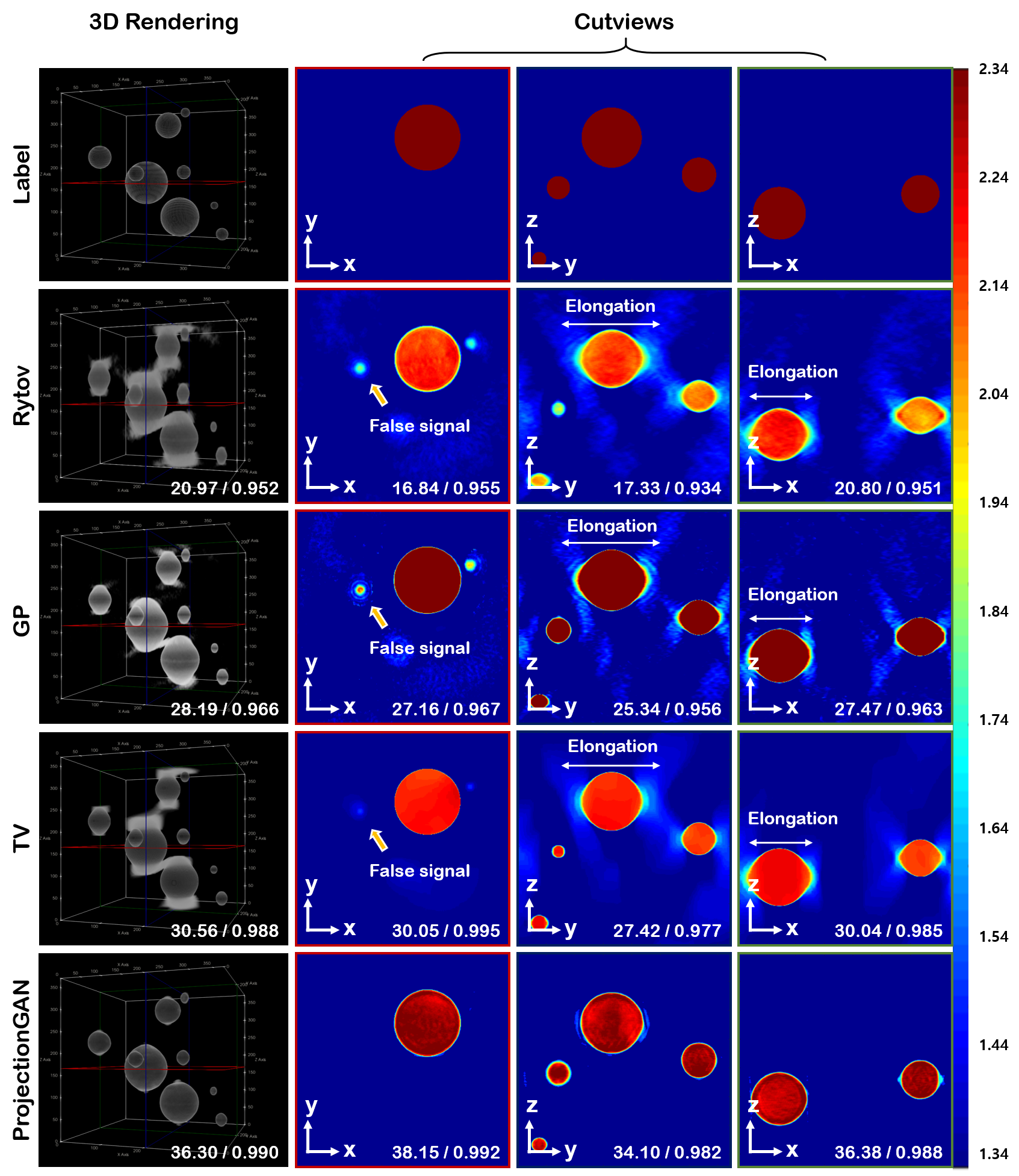}
\caption{Reconstruction of simulated numerical phantom (large). The top row shows ground-truth data, and the following rows depict reconstructions with Rytov, GP, TV, and ProjectionGAN. The leftmost column shows 3D-rendered results of the reconstruction, and the following three columns show x-y, y-z, and x-z planes. \add{Numbers in the bottom right indicate PSNR and SSIM values, respectively. }}
	\label{fig:fig3_numerical_phantom}
\end{figure}

\add{In this section, we provide experimental results of the proposed reconstruction method compared to other approaches. All the designed experiments aim to check the extent to which the algorithms can remove missing cone artifacts. In particular, we anticipate observing reduced elongation along the optical axes and false signals corrupted with noise in the object domain. When visualizing the $k$-space, we expect to observe that the $k$-space is more complete so that the cone-shaped wedge is no longer visible.}

\add{To verify ProjectionGAN both quantitatively and qualitatively with existing reference, a simulation study is presented first in Section~\ref{sec:results_simulation}. Then, we proceed to real ODT measurements to demonstrate that ProjectionGAN can be easily extended to physical acquisitions. The experiment with microbeads in Section~\ref{sec:results_microbead} allows us to perform semianalytic comparison since we have a priori knowledge about the shape of the object and specifically its RI value. Finally, the experiment with biological cells in Section~\ref{sec:results_bio} demonstrates the applicability of ProjectionGAN to complex structures and its generality to diverse sets of measurements.}

\subsection{Numerical Simulation}
\label{sec:results_simulation}

We first performed a numerical simulation study using a $372\times372\times372$ 3D phantom dataset generated using randomly positioned spheres of various sizes. In Fig.~\ref{fig:fig3_numerical_phantom}, we demonstrated that ProjectionGAN accurately reconstructs the ground truth, resolving the elongation issue stemming from the missing cone problem. Moreover, in the axial slice, we can eliminate false signals coming from missing cone artifacts. In contrast, the GP and TV algorithms fall short of the quality compared to ProjectionGAN, where we can clearly see that the algorithm does not fully eliminate the artifacts. \add{The numbers in Fig.~\ref{fig:fig3_numerical_phantom} shows the peak signal-to-noise ratio (PSNR) and structural similarity index (SSIM), respectively. Consistent with the qualitative observation, the metrics also show that ProjectionGAN provides accurate reconstruction, closely matching the ground truth. Other methods largely fall behind ProjectionGAN.}

\begin{figure}[!hbt]
    \centering\includegraphics[width=8.5cm]{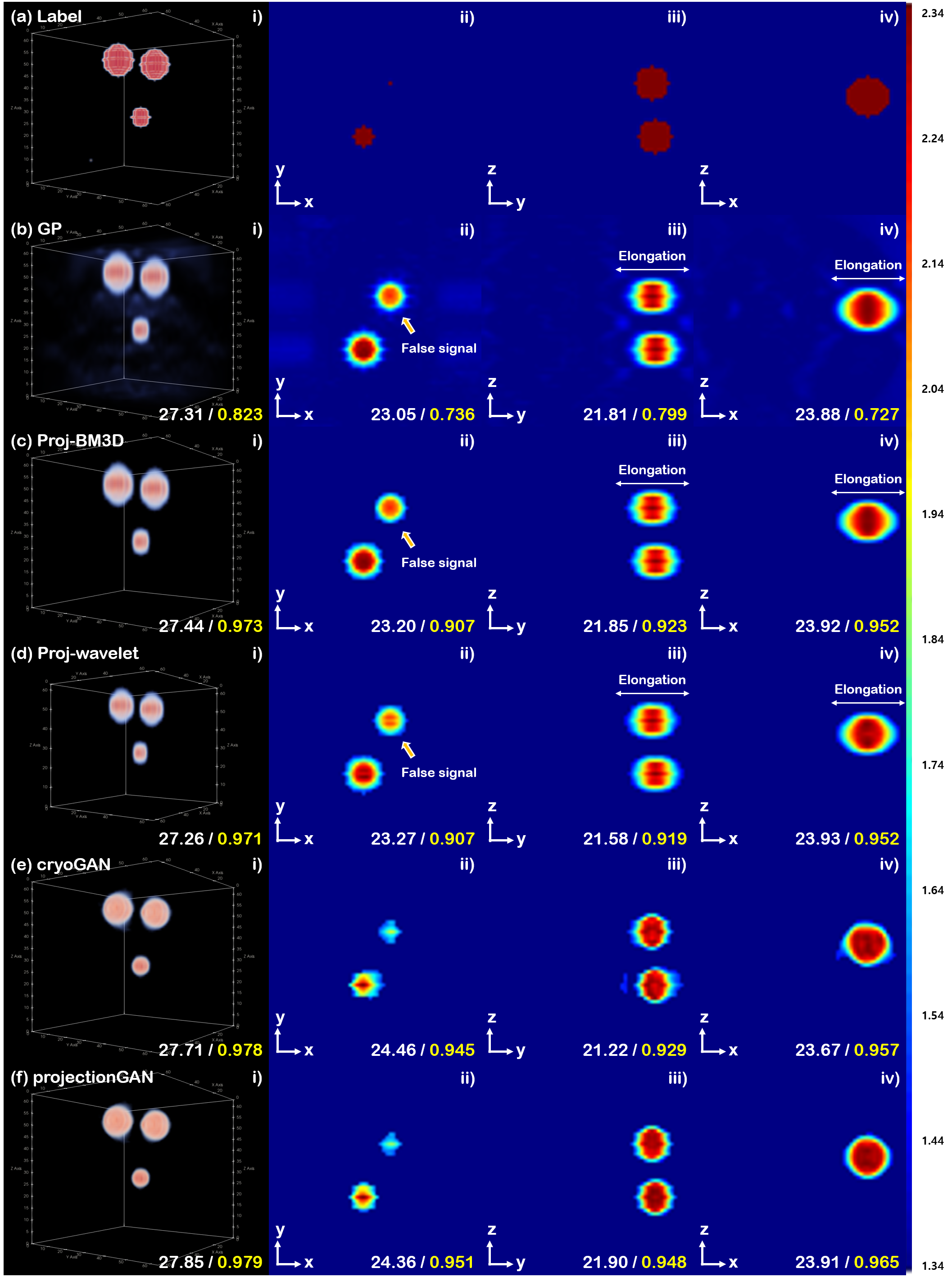}
\caption{Reconstruction of a simulated numerical phantom (small). (a) Label, (b) GP, (c) Proj-BM3D, (d) Proj-wavelet, (e) cryoGAN and (f) ProjectionGAN reconstruction. i) 3D rendered result, ii) x-y plane, iii) y-z plane, iv) x-z plane. \add{Numbers in the bottom right corner of each image correspond to the PSNR and SSIM values, respectively. }}
	\label{fig:cryoGAN_comparison}
\end{figure}

Additionally, we performed a simulation study using the same dataset but of smaller size ($64\times64\times64$) to compare ProjectionGAN with cryoGAN~\cite{gupta2020cryogan} under our computational resources. The results are depicted in Fig.~\ref{fig:cryoGAN_comparison}. \add{Further, we include comparisons with Proj-BM3D and Proj-wav in Fig.~\ref{fig:cryoGAN_comparison}(c)(d)}. Compared to the label data in Fig.~\ref{fig:cryoGAN_comparison}(a), due to the low axial resolution, GP reconstruction in Fig.~\ref{fig:cryoGAN_comparison}(b) shows false signals in the x-y plane and elongation along the optical axes in the y-z and x-z planes. \add{Projection domain denoisers: Proj-BM3D and Proj-wavelet effectively remove background noise apparent in GP but do not ameliorate the optical elongation effect.} In contrast, cryoGAN reconstruction in Fig.~\ref{fig:cryoGAN_comparison}(e) and ProjectionGAN reconstruction in Fig.~\ref{fig:cryoGAN_comparison}(f) both show highly improved results, where false signals are effectively removed and the elongations are suppressed. \add{The quantitative metrics also show the clear superiority of ProjectionGAN over other methods. In particular, the SSIM values are dramatically enhanced by removing the missing cone artifact.} However, it should be noted that cryoGAN is a specimen-specific reconstruction algorithm. Consequently, the time taken even for the $64\times64\times64$ volume was 2 hours. This is a much heavier computational burden than that of ProjectionGAN, which only takes a few seconds.

\subsection{Microbead Measurement}
\label{sec:results_microbead}

\begin{figure}[hbt!]
    \centering\includegraphics[width=8cm]{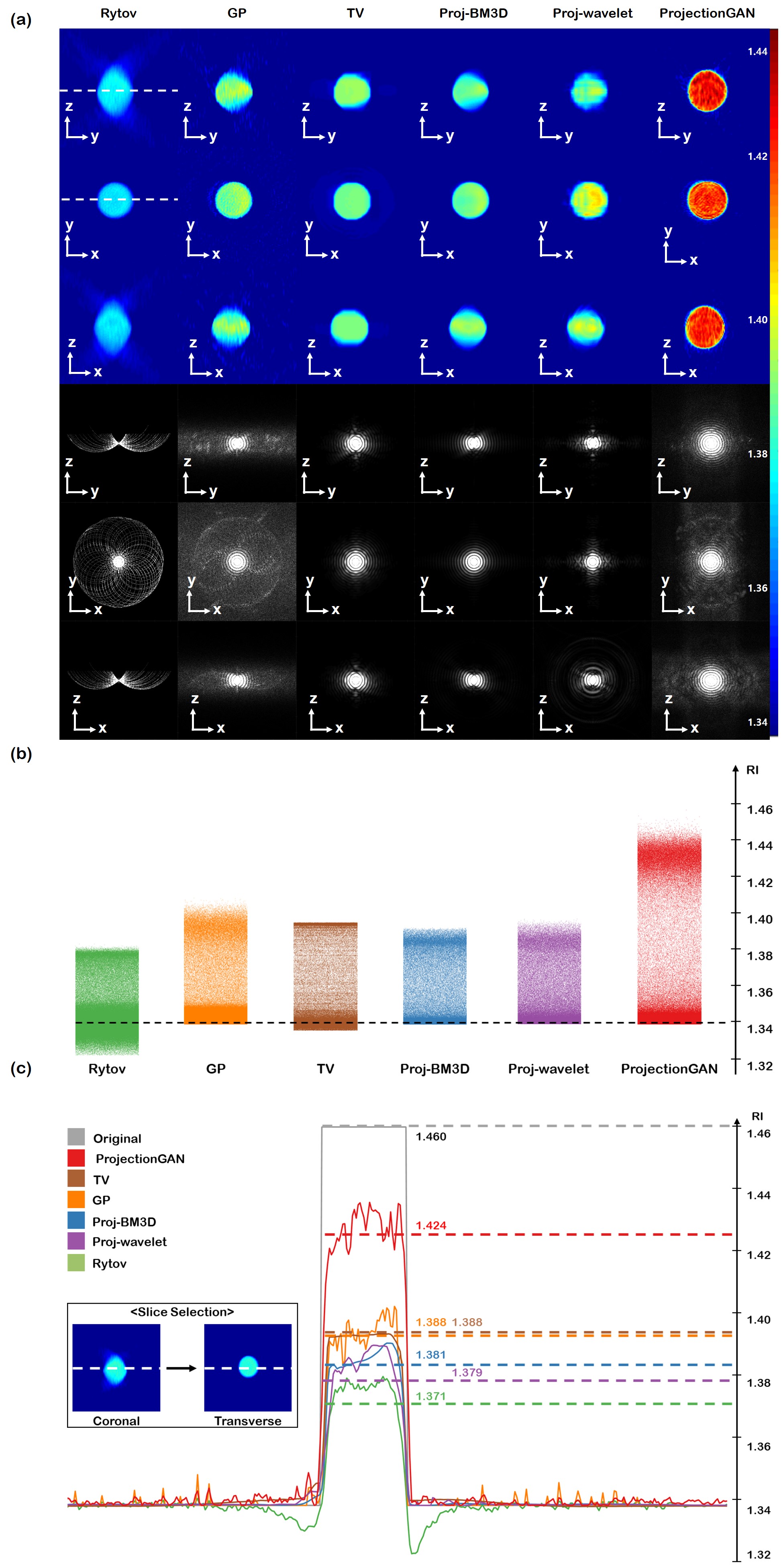}
\caption{Reconstruction of a microbead whose original value is 1.46. (a) Three different slices from different planes (y-z, x-y and x-z planes) are shown for each algorithm. Each slice was taken from the center of the volume. The corresponding visualization of the $k$-space is also given in the lower half. \add{Each column corresponds to reconstructions using the conventional Rytov reconstruction, GP reconstruction, total variation (TV) reconstruction, Proj-BM3D, Proj-wavelet, and the proposed method. (b) Histogram analysis of the RI distribution. The distribution is shifted towards the true RI value with the proposed method compared to other methods. (c) Line profile along the $x$-axis. The mean RI values are improved by the proposed method towards the actual value. }}
	\label{fig:fig2_microbead}
\end{figure}

Furthermore, to verify the efficacy of our method with real ODT measurements, we first tested the network with silica microbeads whose original RI value was approximately 1.46.

In Fig. ~\ref{fig:fig2_microbead} (a), we clearly show that the shape and the RI value are reconstructed with much better accuracy with the use of ProjectionGAN. We compared our method with the conventional Rytov reconstruction, GP reconstruction, TV reconstruction, \add{projection-BM3D, and projection-wavelet thresholding.}
Specifically, we see that with Rytov reconstruction, the original shape of the microbead is not preserved. Moreover, elongation along the optical axes is clearly seen, especially in the y-z and x-z planes. Reconstructions through the GP algorithm resolve the issue of elongation to some extent, but they do not capture the original shape properly. Furthermore, it is evident that the RI values are largely underestimated compared to the actual value due to the missing cone problem in the $k$-space that refers to the 3D Fourier space. The TV reconstruction results show better results: the method produces a more homogeneous structure since it explicitly demands smoothness of the reconstruction. Nonetheless, cutviews seen from the y-z or x-z plane still do not capture the spherical shape properly, not to mention the underestimated RI values. Moreover, missing cone problems are still visible in the $k$-space. \add{Both the projection-BM3D and projection-wavelet thresholding approaches show promise in the elimination of optical elongation artifacts. However, the underestimation of RI values can be clearly seen in Fig.~\ref{fig:fig2_microbead}(b)(c). Blocky artifacts arising from thresholding the wavelet coefficients for Proj-wavelet are also an issue.}

The fourth column in Fig. ~\ref{fig:fig2_microbead} (a) shows reconstructions with the proposed method, where the spherical shape of the microbead is preserved, and underestimated RI values are obtained. Additionally, the $k$-space shows an approximate sinc function, which should be seen if a homogeneous sphere is reconstructed. To inspect the RI distributions in detail, in Fig.~\ref{fig:fig2_microbead}(b), we see the histogram plot of RI values within the 3D data. The intensity of the color bar represents the number of bins in the data, where we see a clear shift towards the actual value with our proposed method compared to the Rytov reconstruction, the GP method, and the TV method. Similar results can be seen in Fig.~\ref{fig:fig2_microbead}(c) with the line profile.

\begin{figure}[hbt!]
    \centering\includegraphics[width=1.0\linewidth]{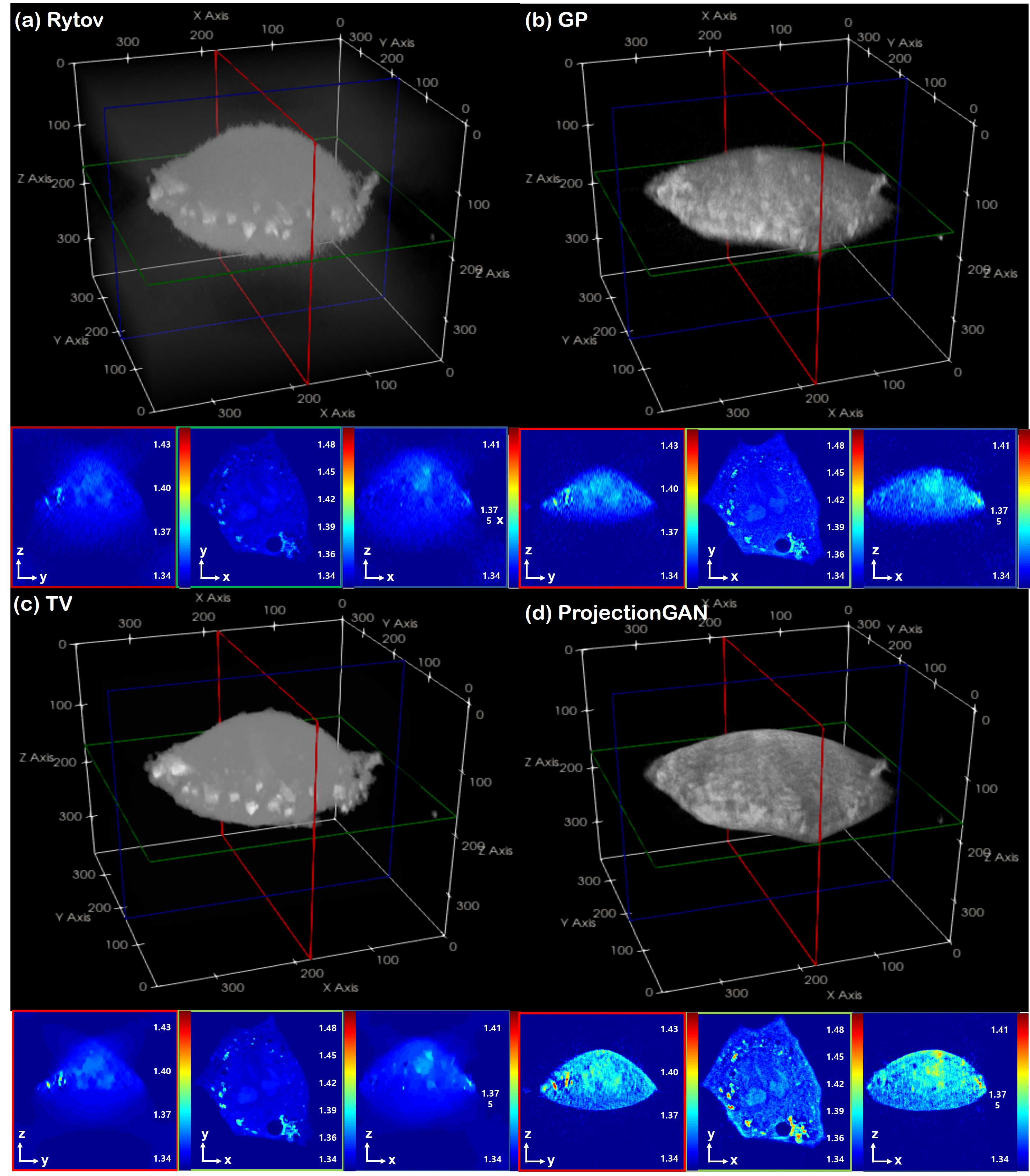}
\caption{3D volume rendering reconstructed with the (a) Rytov, (b) GP, (c) TV, and (d) ProjectionGAN methods. In the view angle shown in the figure, the resolution is very low with the Rytov and TV methods. The GP algorithm has higher resolution but is still below expectations. ProjectionGAN produces high-resolution views at all angles. Cutviews that are covered in red, green, and blue boxes refer to the y-z plane, x-y plane, and x-z plane, respectively. The cutviews also show images with superior resolution, along with higher RI values.}
	\label{fig:3D_rendering}
\end{figure}

\begin{figure}[!hbt]
    \centering
    \includegraphics[width=1.0\linewidth]{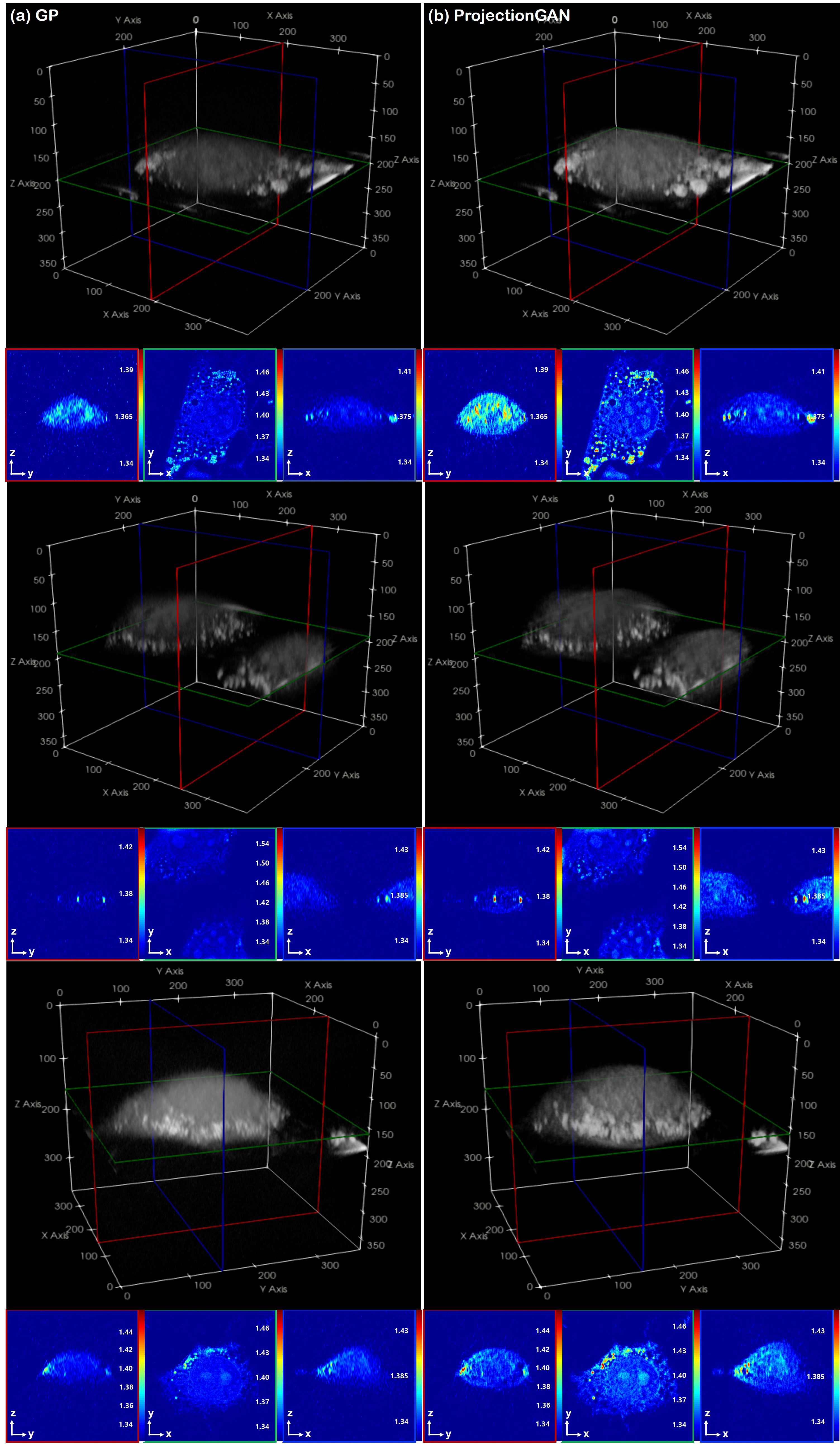}
\caption{3D rendering RI distribution of  cells. Slice views that are marked with red, green, and blue boxes correspond to the y-z, x-y and x-z planes, respectively. (a) GP reconstruction, (b) ProjectionGAN reconstruction.}
    \label{fig:further_results}
\end{figure}

\begin{figure}[!hbt]
    \centering
    \includegraphics[width=1.0\linewidth]{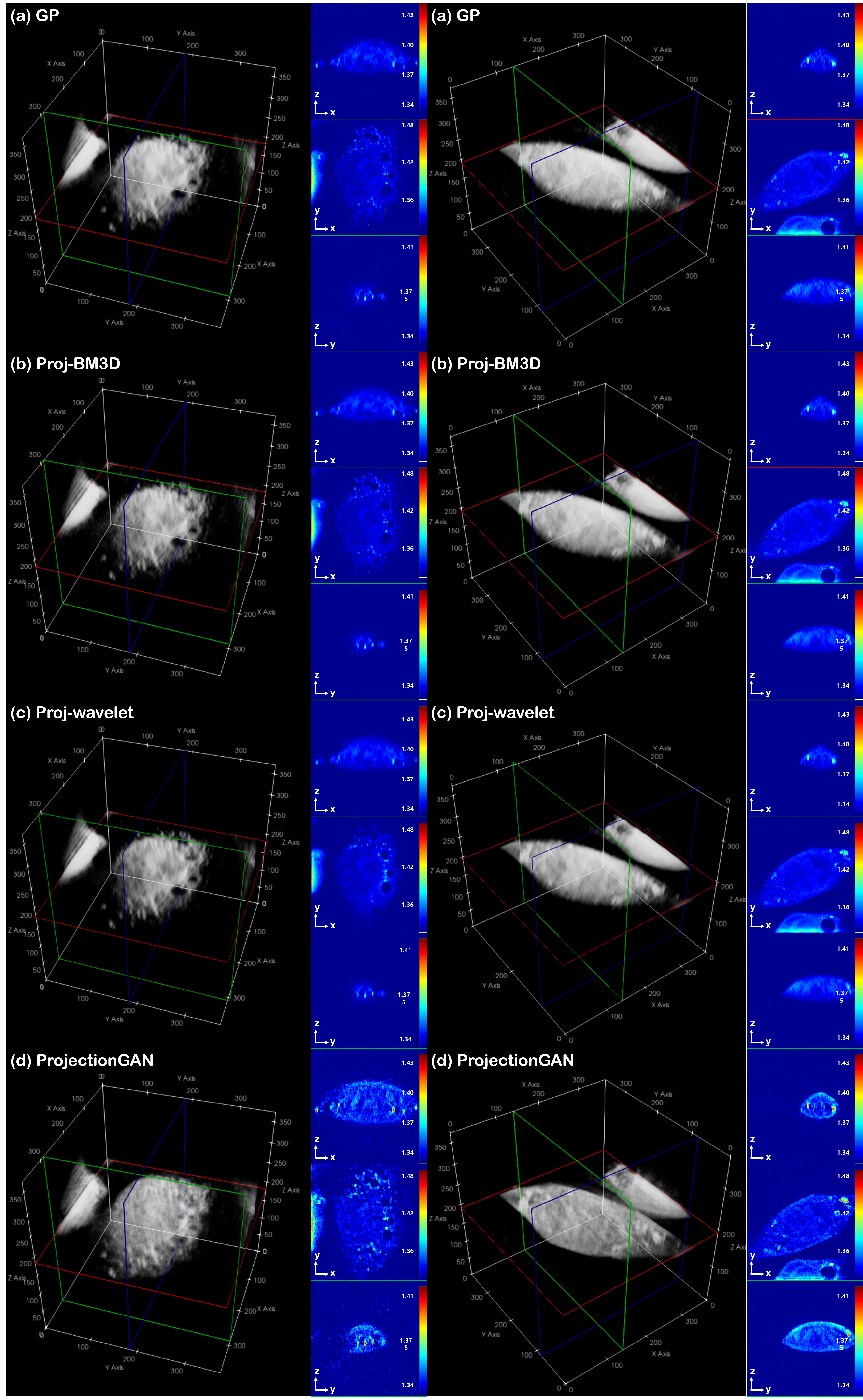}
\caption{\add{Comparison between the GP (a), projection-BM3D (Proj-BM3D) (b), projection-wavelet thresholding (Proj-wavelet) (c), and ProjectionGAN (d) methods}. Slice views that are marked with red, green, and blue boxes correspond to the x-y, y-z, and x-z planes, respectively.}
    \label{fig:projection-BM3D_comparison}
\end{figure}

\begin{figure}[hbt!]
    \centering\includegraphics[width=0.9\linewidth]{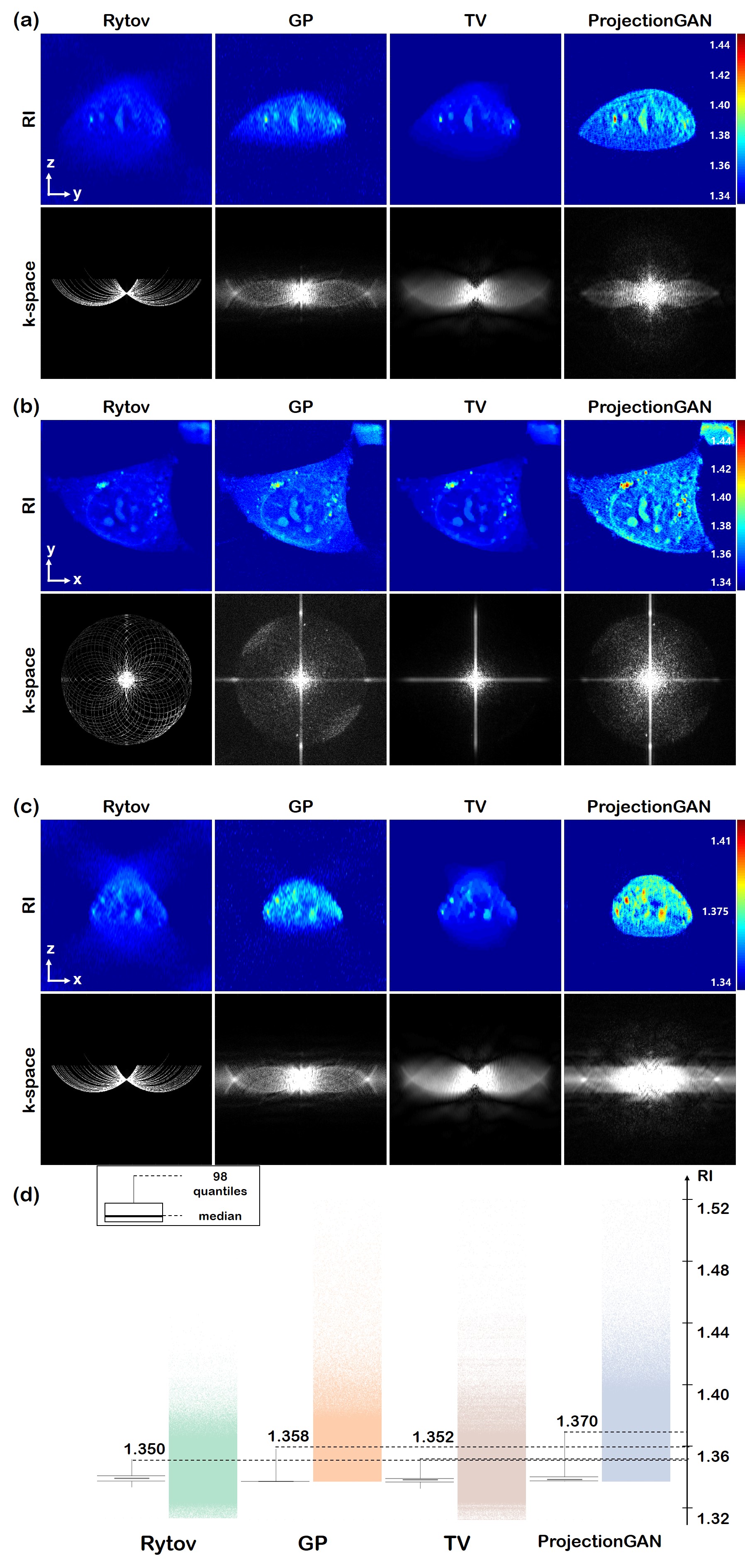}
\caption{Cutviews of 3D RI distribution reconstructed with the Rytov, GP, TV, and ProjectionGAN methods. (a) y-z cutview and the corresponding $k$-space (b) x-y cutview and the corresponding $k$-space (c) x-z cutview, and the corresponding $k$-space. (d) Histogram of the RI distribution of each reconstruction. The number written above the thin dotted line shows the value of the 98\textsuperscript{th} percentile, where we see a shift to higher values with ProjectionGAN.}
	\label{fig:further_results_kspace}
\end{figure}

\subsection{Biological Cells}
\label{sec:results_bio}

Next, we used real biological samples, namely, NIH3T3 cells, to validate our method. In Fig.~\ref{fig:projection_views}, we see severe degradation of image quality as the projection angle deviates from the measurement angle with all the existing methods, i.e, the Rytov, GP, and TV methods. This figure pictorially shows the critical effect of the missing cone problem---the resolution heavily deteriorates in the missing angles. However, using ProjectionGAN, the resolution of projection views is largely enhanced, and we can observe the cell clearly at all angles.

Using the improved projection data in Fig.~\ref{fig:projection_views}, we performed three-dimensional reconstruction using filtered backprojection for in-depth analysis of the reconstructed RI distribution. In Fig.~\ref{fig:3D_rendering}, we provide 3D rendering of the reconstructed NIH3T3 cell distribution and its cut-views. Here, in the view presented in Fig.~\ref{fig:3D_rendering}, we see that the resolution in the reconstructions with the Rytov and TV approaches are far below satisfactory, where we cannot clearly see the configuration of cell organelles. In particular, TV reconstruction in Fig.~\ref{fig:3D_rendering}(c) shows a cartoon-like appearance, which does not capture the true texture of the cell. With GP reconstruction, as shown in Fig.~\ref{fig:3D_rendering}(b), we achieve higher resolution with more details preserved. Nevertheless, the resolution in the y-z and x-z planes is still below the standards.
Specifically, when we visualize the cut-view RI distribution, the images in the y-z and x-z planes have very low resolution, whereas images reconstructed through ProjectionGAN show superior results in all three planes. Here, we again confirm that the problem of underestimated RI values is resolved. The same superiority can be seen in the reconstruction results of other types of cells, as depicted in Fig.~\ref{fig:further_results}

Furthermore, we  present the comparison with projection-BM3D approach, \add{and projection-wavelet thresholding approach,} as stated in \ref{sec:projection-BM3D}. The results are summarized in Fig.~\ref{fig:projection-BM3D_comparison}. From the results, we observe that applying BM3D \add{or wavelet thresholding} to the projections reduces the overall noise apparent in the GP reconstructions. Unfortunately, unlike ProjectionGAN, where by learning the distribution mapping $\Yc_{\Omega^c} \mapsto \Yc_{\Omega}$, the network learns not only to denoise but also to capture the structure and eliminate the optical elongation, BM3D \add{or wavelet thresholding} merely succeeds at eliminating background noise. Here, we verify that using CycleGAN as the projection domain enhancer is a critical part of the design of ProjectionGAN.

Finally, we provide slice views from the Rytov, GP, TV and ProjectionGAN methods in each plane along with the $k$-space of each corresponding slice in Fig.~\ref{fig:further_results_kspace}(a)-(c).
With Rytov reconstruction, the image support in the y-z and x-z planes is vague, and the resolution is highly degraded. The GP method has higher resolution than Rytov reconstruction, but the cell support is vague. TV reconstruction shows cartoon-like artifacts, which do not properly capture the high-frequency details of the cell structure. On the other hand, with our method, we are able to achieve high resolution on all three axes with clear boundaries. The enhancement of the RI values is also remarkable.

The missing cone problem can be best characterized when the $k$-space is visualized. More specifically, due to the Fourier diffraction theorem, field retrieval is performed in 3D $k$-space, where each hologram corresponds to the half-sphere of the $k$-space. Since 49 holograms were acquired to reconstruct a single RI distribution in our case, in Rytov reconstruction in Fig. \ref{fig:further_results_kspace}, 49 curves are apparent, regardless of the view plane. The $k$-space of reconstructions with the GP method solves the missing cone problem by utilizing a nonnegativity constraint, and thus, when seen in the x-y plane, it seems that the missing cone problem is resolved. Nonetheless, when seen from the y-z or x-z plane, we still observe the missing parts of the $k$-space. With TV reconstruction, a similar phenomenon was observed in the spectrum of TV reconstruction, which resembles a blurred version of Rytov or GP reconstruction.
However, our new method ProjectionGAN resolves the problem completely regardless of the viewing plane. Note that the half-spherical curves that were apparent in the Rytov, GP and TV reconstruction are no longer visible in the $k$-space reconstructed with the proposed method.

\section{Conclusion}
\label{sec:conclusion}

In this paper, we proposed a novel ProjectionGAN that resolves the missing cone problems in ODT using a novel unsupervised learning approach. Specifically, an initial GP reconstruction was used to generate the projection views at the missing view angles, which were improved by a novel ProjectionGAN algorithm.
Our ProjectionGAN algorithm relies on the optimal transport-driven CycleGAN that learns to transport projection distributions from missing views to those of high-quality projections. After the projections are refined, the 3D RI distribution can be reconstructed through a simple filtered backprojection.
The experimental results confirmed that the characteristics of the scattering potential from our method are largely superior to those of conventional reconstructions.

\section*{Acknowledgment}
This work was supported by the Korea Medical Device Development Fund grant funded by the Korean government (the Ministry of Science and ICT, the Ministry of Trade, Industry and Energy, the Ministry of Health \& Welfare, the Ministry of Food and Drug Safety) (Project Number: 1711137899, KMDF\_PR\_20200901\_0015). This research was funded by the National Research Foundation (NRF) of Korea grant NRF-2020R1A2B5B03001980.

\ifCLASSOPTIONcaptionsoff
\newpage
\fi

\bibliographystyle{IEEEtran}
\bibliography{reference}
\end{document}